\documentclass[11pt]{article}
\usepackage{moriond_mod}
\usepackage{epsfig}  
\usepackage{amssymb}

\newcommand{\rpv}{$R_p\hspace{-2.3ex}/\hspace{2.3ex}$}

\begin{document}

%\vspace*{4cm}
\vspace*{3cm}

\title{
Searches beyond the Standard Model at HERA}

\author{
Andr\'e Sch\"oning \footnote{Talk given at the XXXVI\,th Rencontres de Moriond ``Electroweak Interactions and Unified Theories'', 10-17 March 2001.}}

%\date{talk presented in XXXVIth Rencontres de Moriond Electroweak Interactions
% and Unified Theories, 10-17 March 2001}

\address{Institute for Particle Physics, ETH Z\"urich, CH-8093 Z\"urich, Switzerland\\
(on behalf of the H1 and ZEUS collaborations)}

\maketitle

\abstracts{
At HERA, new physics processes beyond the SM are probed in $e^\pm$-proton 
collisions at a center of mass energy of $300-318\,\rm GeV$. 
Recent results on searches obtained 
by the H1 and ZEUS ex\-pe\-ri\-ments are presented with 
an emphasis on data taken in $e^-p$ collisions in the years
1998/99 and in $e^+p$ collisions in the years
1999 and 2000.}

%\newpage

\section*{Introduction}

At the HERA collider, electrons (positrons) and protons are collided at a
center of mass energy of about 
$\sqrt{s}=318\,\rm GeV$ ($300\,\rm GeV$ before 1998), directly probing  
new physics in
$eq$ interactions at the highest energies. Since 1994, integrated luminosities
of about ${\cal L}=115\,\rm pb^{-1}$ in $e^+p$ and 
${\cal L}=15\,\rm pb^{-1}$ in $e^-p$ scattering have been collected by the
two experiments H1 and ZEUS.
The most recent data collected in $e^+p$ scattering
in the most successful year 2000, corresponding to
about ${\cal L}=60\,\rm pb^{-1}$,
increased significantly 
the sensitivity to new processes. First results are presented here.

\section*{Indirect Searches}
\begin{figure}[tb] \unitlength 1mm
% \begin{center}
   \begin{picture}(55,80)
    \put(0,-2){\epsfig{file=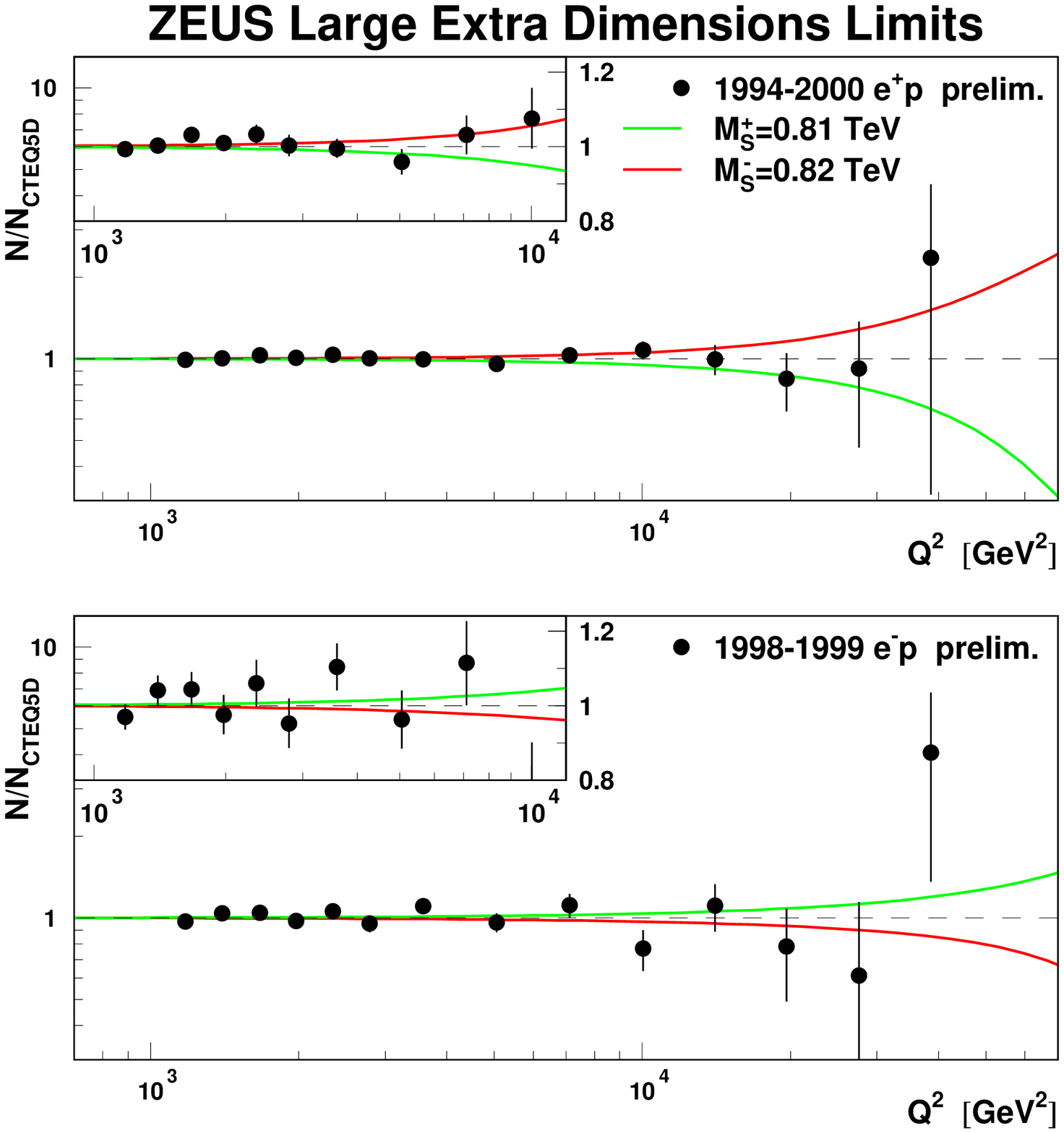,bbllx=20,bblly=300,bburx=560,bbury=840,clip=2,width=0.50\textwidth}}
    \put(87,-4){\epsfig{file=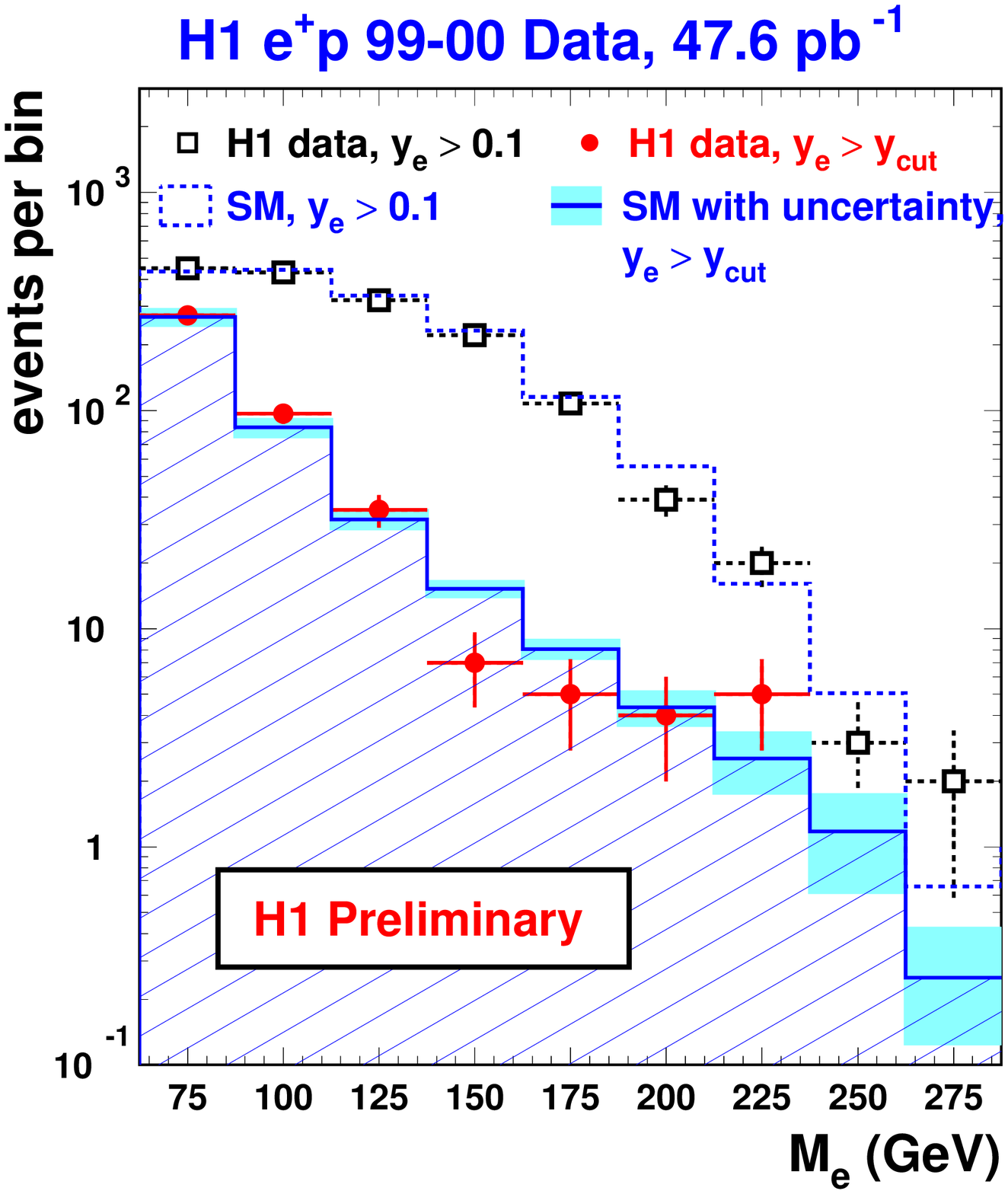,width=0.5\textwidth}}
    \put(0,80){\bf (a)}
    \put(85,80){\bf (b)}
   \end{picture}
% \end{center}
\caption{(a) Ratio of the observed number of events over the expectation in Neutral Current DIS
as a function of $Q^2$ as measured in $e^+ p$ and $e^- p$ scattering by ZEUS.
The fitted curves show the 95\% CL exclusion limits on large extra dimensions
for constructive and destructive interference with the SM process.
(b) Distribution of the $e - \rm  jet$ 
invariant mass as reconstructed from the scattered positron for all
events ({\sl open squares}) and after applying an optimised angular cut
({\sl full circles}) to enhance the discrimination between leptoquarks and 
SM processes. The data were taken by H1 in the most recent 1999-2000 $e^+p$ run.
}
\label{pl:dis}
\end{figure}
Measuring the deep inelastic scattering process (DIS)
$e^\pm p \rightarrow e^\pm X $ at the highest momentum transfers ($Q^2$) 
% see fig.~\ref{pl:dis}~(a),
gives indirect access
to new particles such as heavy bosons, 
leptoquarks and composite particles or to quantum
gravity effects above the center of mass energy of HERA. 
%The neutral current cross-section as a function of the 
%squared four momentum transfer $Q^2$ is shown in fig.~\ref{pl:dis}(a)
%for both $e^+p$ and $e^-p$ scattering. 
No deviation from the SM expectation was found in the most recent data
\cite{ref:nc_h1,ref:nc_zeus}. 
Therefore limits on new physics processes have been set using
the formalism of four-fermion (point-like) contact interactions.
%Limits are derived by fixing the SM couplings and by fitting additional new
%vector-type couplings to the data:
% \begin{equation}
% L_V \ = \ \sum_{a,b = L,R} { \eta_{ab}^q} \,
%         (\bar{e}_a \gamma^\mu e_a) \ (\bar{q}_b \gamma_\mu q_b) \ \ \ \ \ ,
% \end{equation}
Limits are derived by fixing the SM couplings and by fitting additional new
vector-type couplings $
L_V \ = \ \sum_{a,b = L,R} \  { \eta_{ab}^q} \,
(\bar{e}_a \gamma^\mu e_a) \ (\bar{q}_b \gamma_\mu q_b)
$ to the data,
where $\eta_{ab}^q  \ = \ \epsilon \, \frac{g^2}{(\Lambda_{ab}^q)^2}$
are model-dependent coefficients of the new process,
$g$ is the coupling constant, $\Lambda_{ab}^q$ is the effective mass scale
and $\epsilon=\pm 1$ is a parameter determining the interference with the SM.
Fits were performed simultaneously to the $e^+p$ and $e^-p$ data and
95\% confidence limits (CL) were derived. 
Using almost full data statistics, H1 
excludes compositeness scales~\cite{ref:osaka_ci} up to about 9~TeV 
assuming $g^2=4\pi$.

Similarly, limits can be set on models with 
Large Extra Dimensions (LED) \cite{ref:led}, see fig.~\ref{pl:dis}~(a),
in which SM particles are bound
to a (3+1)-dim. space while gravitons live in a world with $\delta \ge 2$ extra 
compactified dimensions. This model solves the hierarchy problem
by bringing down gravitation to the TeV scale, i.e. the electroweak scale.
Gravitons are expected to appear in the (3+1)-dim. world as towers of 
Kaluza-Klein modes and gravitation is actually a strong force 
at short distances below 
%the compactification radius $r<R$. 
the size of the compactified extra dimensions.
%Because in experiments 
Gravitation has been tested directly only down to the milli-meter scale,
so it is attractive to test the LED model at HERA and other collider 
facilities 
where compactification at the $\rm TeV$ scale can be tested.
%interactions at distances down to $10^{-16}\,\rm cm$ are probed.

95\% CL limits have been derived by fitting the SM Lagrangian with an 
additional term $\propto \epsilon/M_s^4$,
which accounts for the graviton exchange with $M_s$ being an 
effective compactification mass scale.
95\% CL limits were set by the 
H1 experiment~\cite{ref:osaka_ci}:
$M_s^+>0.6\,\rm TeV$ and $M_s^->0.9\,\rm TeV$
and by the ZEUS experiment: $M_s^\pm>0.8\,\rm TeV$.
%, where the index $\pm$ gives
%the interference with the SM process.

\section*{Direct Searches}
\begin{figure}[thb] \unitlength 1mm
   \begin{picture}(70,65)
    \put(-3,6){\epsfig{file=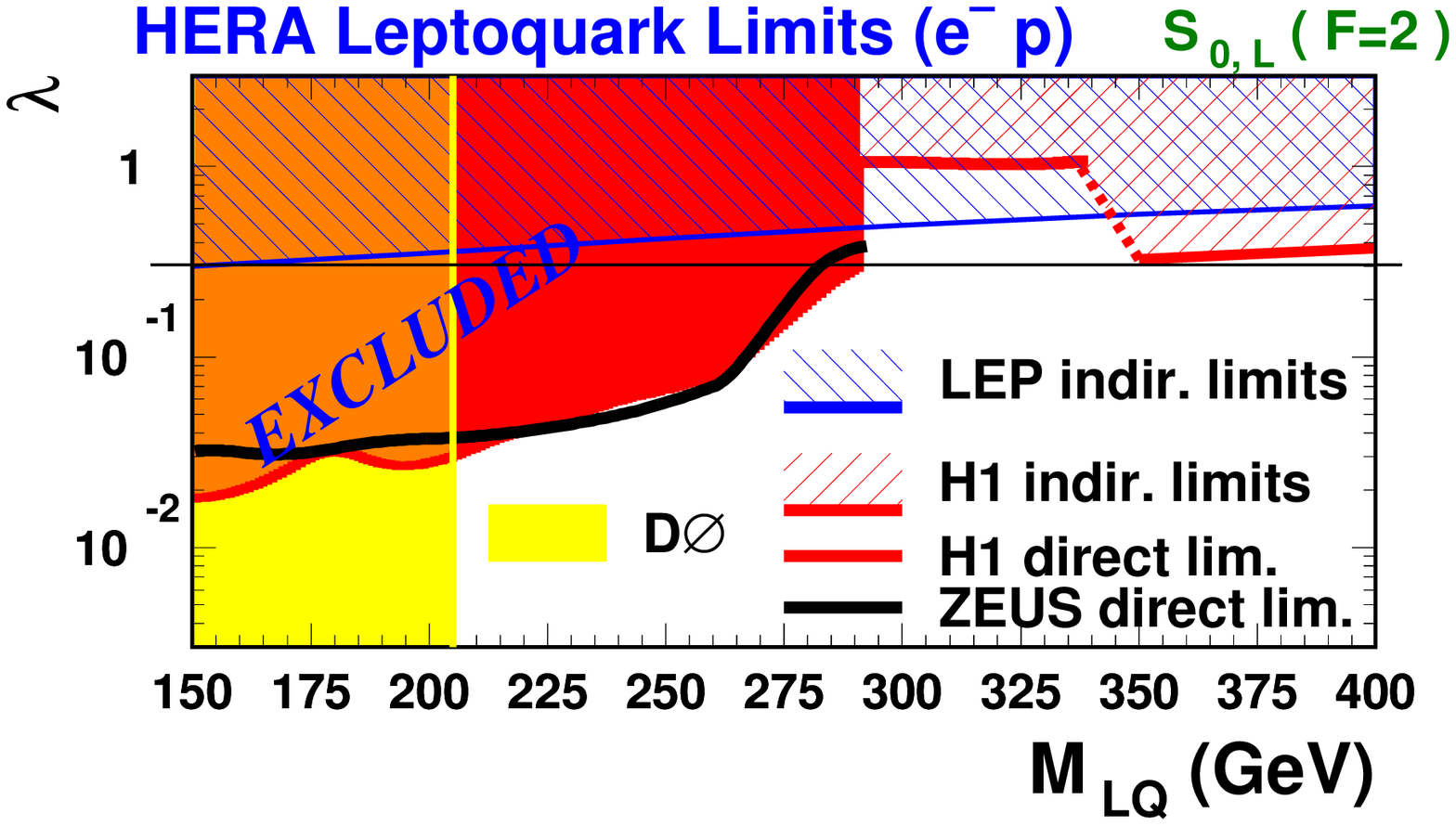,width=0.58\textwidth}}
    \put(90,-4){\epsfig{file=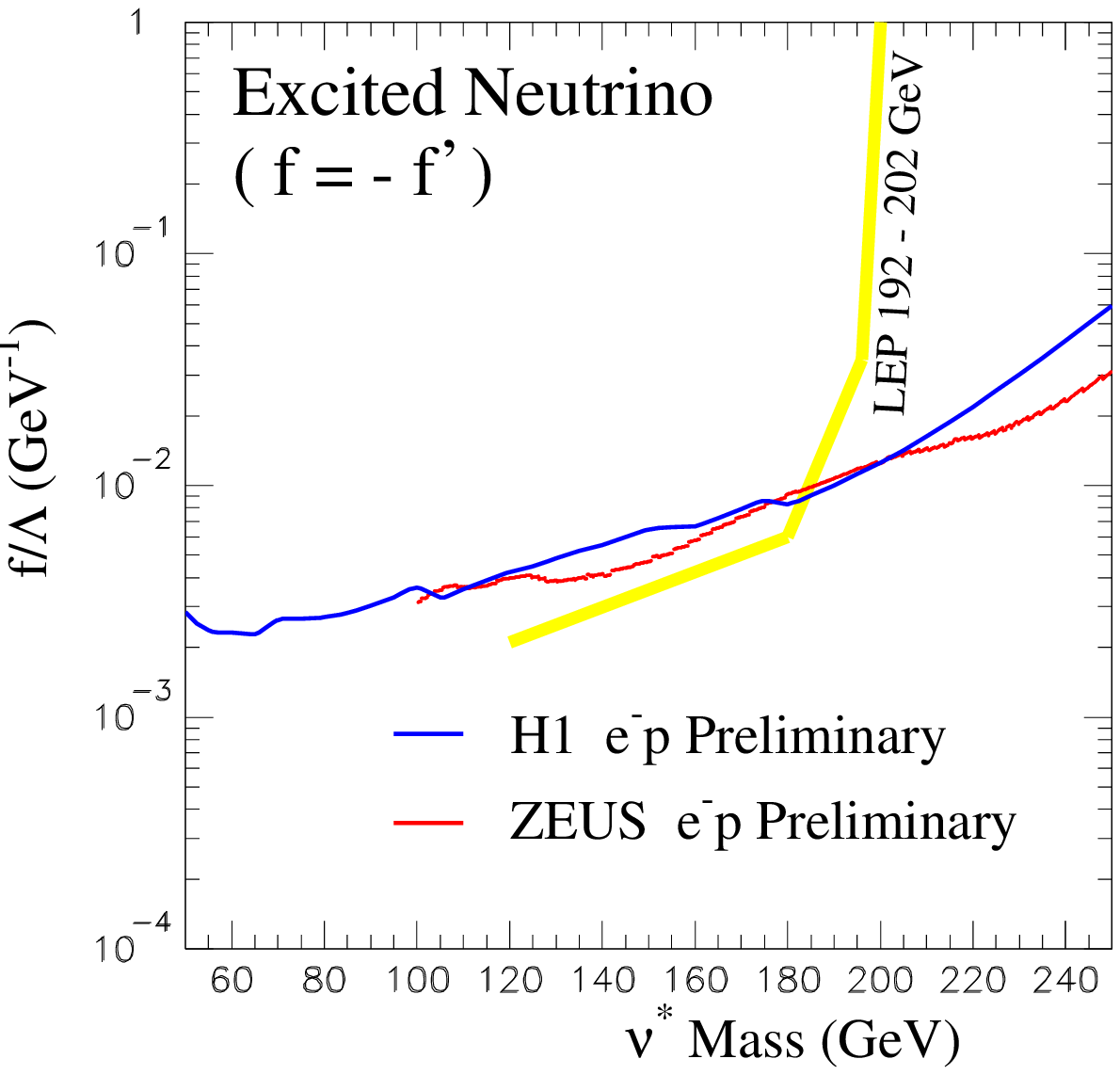,width=0.49\textwidth}}
    \put(0,60){\bf (a)}
    \put(85,60){\bf (b)}
   \end{picture}
\caption{(a) Exclusion region of $\lambda$ as a function of the $S_{0,L}$ 
leptoquark mass. For $M_{LQ}<\sqrt{s}$ the limits from direct searches performed
by H1 and by ZEUS are shown.
For $M_{LQ}>\sqrt{s}$ the limit derived by the indirect search from H1
is shown. 
For comparison, the limits obtained by 
LEP and D0 are also shown.
(b) Exclusion limits on 
$\rm f/\Lambda$ for the excited neutrino search as a function of the $\nu^*$ mass. 
The limits are shown for the coupling assumption $\rm f=-\rm f^\prime$ as measured
by H1 and ZEUS. 
For comparison, the limit obtained by DELPHI~{\protect \cite{ref:exc_delphi}} is also shown.
}
\label{pl:lq+exc}
\end{figure}
%In contrast to indirect searches 
The existence of leptoquarks, excited
fermions and scalar quarks, which are predicted by Grand Unification (GUT), compositeness and supersymmetry (SUSY) models,
can be probed directly by looking
for $eq$-resonances or characteristic decays of new particles.
All these particles would be dominantly singly produced at HERA and
the cross-section depends therefore quadratically on 
the coupling of the new state to SM particles.

The excess observed in neutral current DIS $e^+p$ data 1994-1997 
at positron-quark invariant masses at about $200\,\rm GeV$ by H1
\cite{ref:lq97_h1} was not confirmed by the recent data taken in 1999-2000
for which the invariant mass distribution is shown in 
fig.~\ref{pl:dis}(b)~\cite{ref:lq_h1_eplus}.
A similar excess observed by ZEUS~\cite{ref:dis97_zeus} for masses 
above $220\,\rm GeV$ was also not confirmed by the most recent 
data~\cite{ref:nc_zeus}.

Different types of leptoquarks (LQ) 
can be probed at HERA and results are presented
here in the framework of the BRW model \cite{ref:brw}.
The seven $F=0$ scalar and vector LQs, 
which couple simultaneously to one fermion and one antifermion, 
are better probed in $e^+p$ collisions
due to the large valence quark densities of the proton.
Similarly, the seven $|F|=2$ scalar and vector LQs, 
which couple to two fermions or two antifermions,
are better probed in $e^-p$ collisions.

Searches for all types of LQs have been performed
and no deviation from the SM was found.
Results for a $S_{0,L}$ LQ with fermion number
$F=2$ are shown in fig.~\ref{pl:lq+exc}(a) 
using the BRW model where 
the $S_{0,L}$ LQ decays equally into $eq$ and $\nu q$.
95\% CL limits on the Yukawa coupling $\lambda$ 
%at the LQ-$e$-$q$ vertex
have been derived from the direct searches performed 
by H1~\cite{ref:lq_h1_emin} and ZEUS~\cite{ref:lq_zeus}
as a function of the LQ mass.
For masses above the collider energy, 
the result taken from the indirect contact interaction search
obtained by H1~\cite{ref:lqind_h1} is shown. 
For comparison, the indirect limit obtained by LEP
% {\cite{ref:lq_LEP}} 
and the direct limit from D0, 
which is independent of the Yukawa coupling, are also shown.
% {\cite{ref:lq_D0}}.

\begin{figure}[thb] \unitlength 1mm
% \begin{center}
   \begin{picture}(70,65)
    \put(10,-2){\epsfig{file=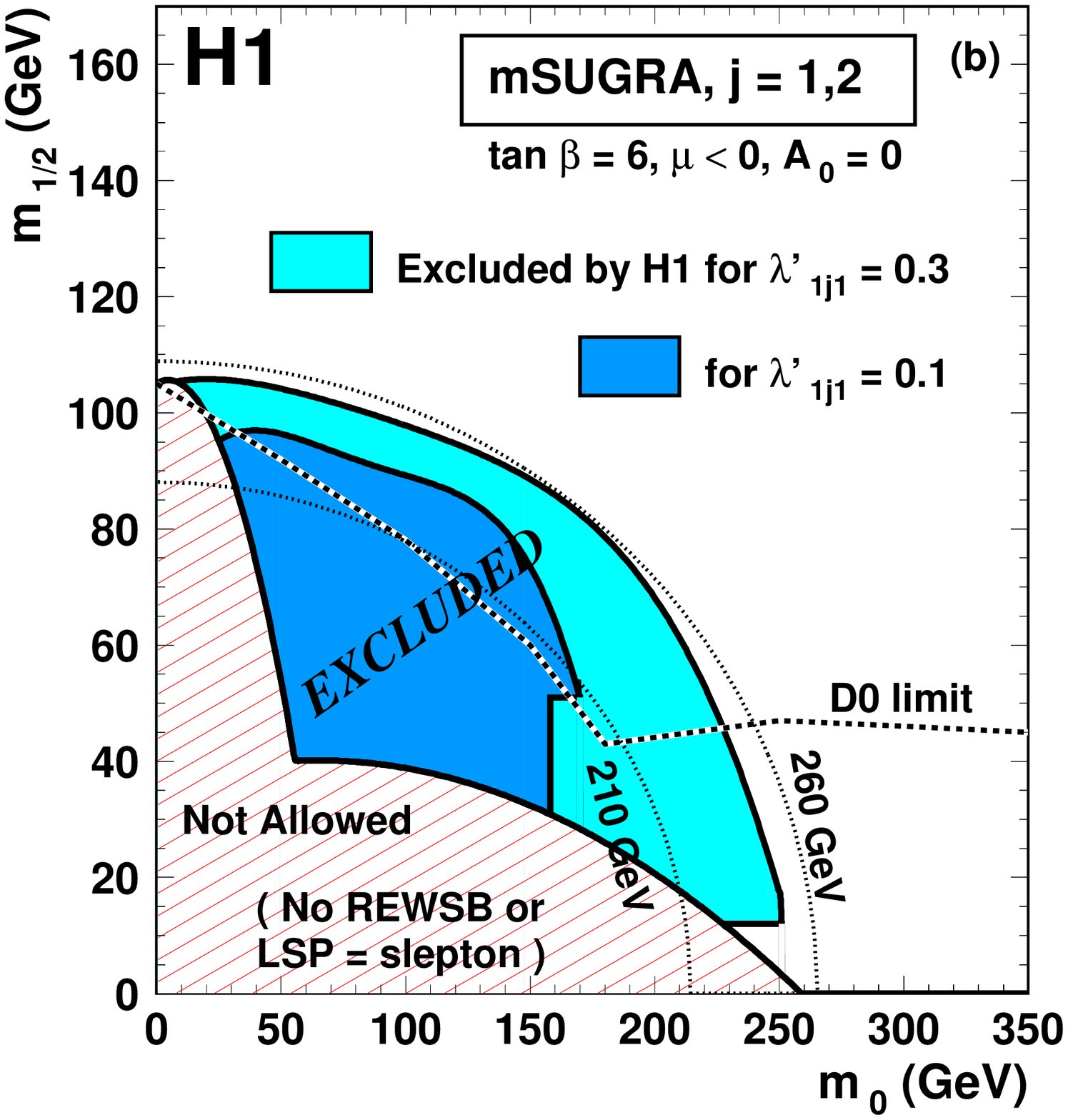,width=0.42\textwidth}}
    \put(0,60){\bf (a)}
    \put(90,-8){\epsfig{file=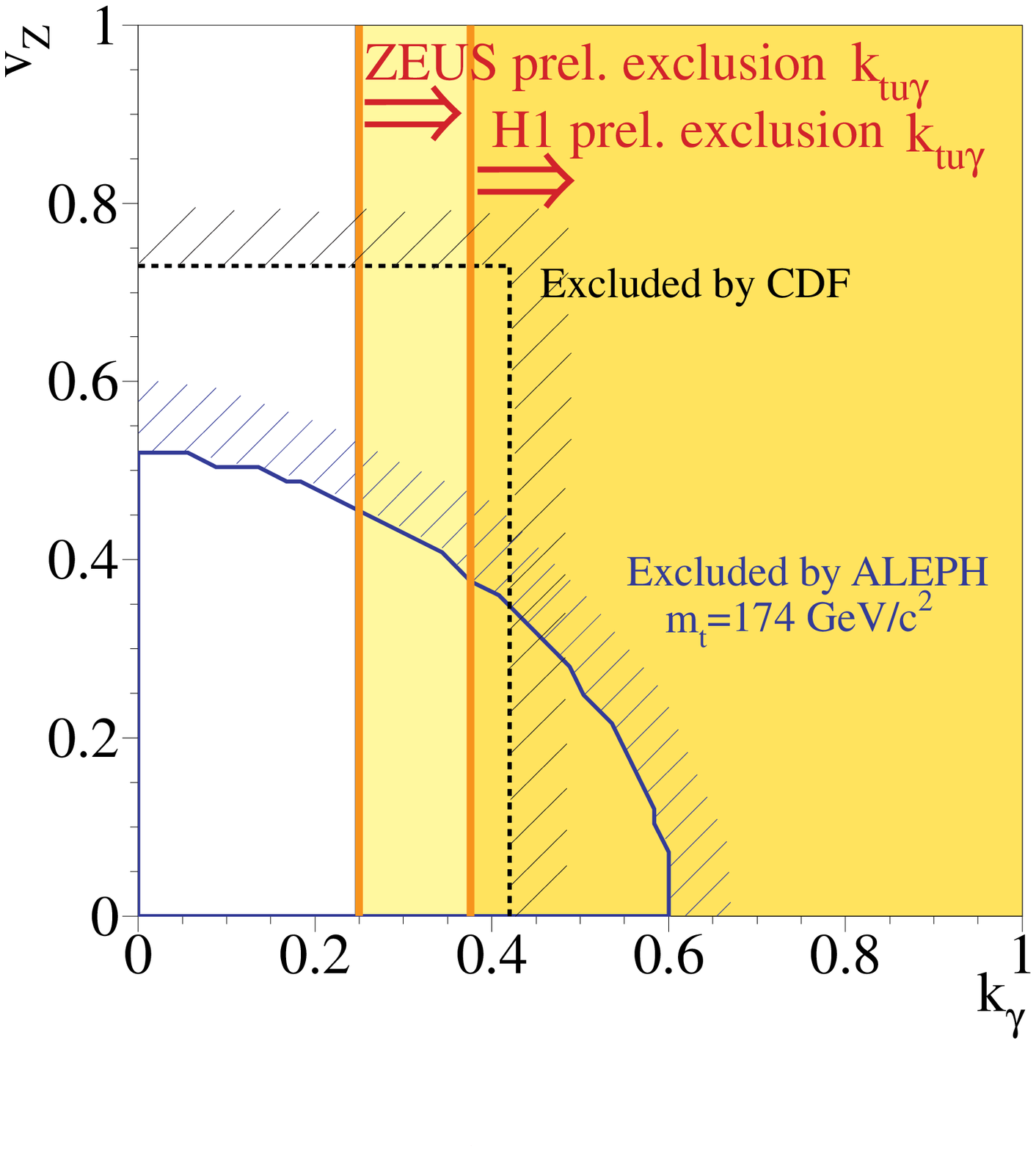,width=0.50\textwidth}}
    \put(85,60){\bf (b)}
   \end{picture}
% \end{center}
\caption{(a) 
Domain of the plane ($m_0,m_{1/2}$) excluded for $\tan \beta = 6$
for a \rpv coupling $\lambda^\prime_{1j1} = 0.1,0.3 $ ($j=1,2$).
The region below the dashed curve is excluded by the D0 experiment
and does not depend on the Yukawa coupling. Also shown are two isolines
for the mass of the $\tilde{u}^j_L$ as dotted curve.
(b) Excluded regions in the plane $v_Z$ versus $\kappa_\gamma$. 
H1 and ZEUS set limits on the anomalous coupling $\kappa_{tu \gamma}$.
For comparison, the exclusion limits obtained by ALEPH 
%{\protect \cite{ref:aleph_ano}} 
and by CDF 
%{\protect \cite{ref:cdf_ano}}
are shown.
}
\label{pl:squark+ano}
\end{figure}

Searches for excited states of fermions like $e^*$, $\nu^*$ and $q^*$ have been 
made by H1~\cite{ref:exc_h1} and ZEUS~\cite{ref:exc_zeus} 
in all final state topologies resulting from the decay of a excited
fermion into an electroweak boson.
No significant deviation from the SM expectation
was found in any of the decay channels.
In general, both the single $f^*$ production and the 
branching ratios of $f^*$ depend on the gauge group weights $\rm f^\prime$, 
$\rm f$ and $\rm f_s$ which describe the relative couplings 
of an excited fermion to the gauge groups 
$U(1)_Y$, $SU(2)_L$ and $SU(3)_c$~\cite{ref:exc}.
Therefore all possible decay channels were combined for certain assumptions
on the relative gauge group weights and limits were
set on the ratio of the gauge group weight over compositeness scale, 
$\rm f/\Lambda$.
As an example, fig.~\ref{pl:lq+exc}(b) shows the HERA limits on $\nu^*$ production
assuming $\rm f = - \rm f^\prime$ for which photonic decays are allowed. 
The limits extend beyond those obtained at LEP to $\nu^*$ masses larger
than $200\,\rm GeV$. Similar limits were set on the production of $e^*$ 
and $q^*$.

Scalar quarks as predicted by SUSY 
can be produced directly via $ep \rightarrow \tilde{q} X$ at HERA 
if \rpv is explicitly broken by the term $\lambda^\prime_{ijk} L_i Q_j \bar{D}_k$
\cite{ref:dreiner}.
%Depending on the parameters of the MSSM model
Scalar quarks can decay either directly into lepton and quark
or via gauge couplings into quark and gaugino followed by a \rpv SUSY decay. 
Since in these particular decay modes Majorana particles 
may be involved, the final state is 
expected to contain electrons (or positrons)
explicitly violating lepton number conservation.
Therefore these decay modes are
almost background free. Searches have been performed by H1~\cite{ref:rpsusy_h1}
and ZEUS~\cite{ref:rpsusy_zeus} which
essentially cover 100\% of all decay topologies.
No significant deviation from the SM expectation was found. 
By making SUSY parameter scans limits on $\lambda$ were set in an
unconstrained MSSM model by both experiments
where the scalar quark mass is treated as a free parameter. 
Scalar quark masses up to about $260\,\rm  GeV$ have been excluded
for all generations for 
$\lambda^\prime_{\rm Yukawa} =\sqrt{4 \pi \alpha_{em}} \approx 0.3$.
The limits were found to be almost independent of the MSSM parameters.
Similar limits were set by H1 for the mSUGRA model.
As an example, 
fig.~\ref{pl:squark+ano}~(a) shows the H1 result 
in the plane ($m_0,m_{1/2}$) for a first or second generation squark. 

\section*{Isolated Lepton Events}
Excitement was caused by the observation of an anomalously large
number of isolated high-energy lepton events with missing transverse 
momentum by the H1 experiment~\cite{ref:isol}.
The ZEUS experiment has presented a similar analysis of isolated lepton events
including all events taken from 1994-2000. 
The result, after applying two different cuts on the transverse momentum of the
hadronic final state ($p_t^X$), is shown in the following table in comparison to
the SM expectation, which is mainly $W$ production.
\vskip 0.3cm
        \begin{tabular}{|c|c|cr|cr|} \hline
\bf Preliminary     &   & \multicolumn{2}{c|}{\bf Electrons} & \multicolumn{2}{c|}{\bf Muons} \\  
\bf Results     &   $p_t^X$ cut   & \multicolumn{2}{c|}{\small Obs./expected (W)} &  \multicolumn{2}{c|}{\small Obs./expected (W)} \\ \hline \hline

\bf H1 1994-2000 &         $ p_t^X > 25$ GeV & ${3}/1.05\pm0.27$ & $(0.83)$ & ${6}/1.21\pm 0.32$ & $(1.01)$ \\ 
% \hline
\bf $\bf e^+ p$ (82 pb$^{-1}$) &         $ p_t^X > 40$ GeV & ${2}/0.33\pm0.10$ & $(0.31)$ & ${4}/0.46\pm 0.13$ & $(0.43)$ \\ \hline \hline
\bf ZEUS 1994-2000  &         $ p_t^X > 25$ GeV & ${1}/1.14\pm0.06$ & $(1.10)$ & ${1}/1.29\pm 0.16$ & $(0.95)$ \\ 
% \hline
\bf $\bf e^\pm p$ (130 pb$^{-1}$) &         $ p_t^X > 40$ GeV & ${0}/0.46\pm0.03$ & $(0.46)$ & ${0}/0.50\pm 0.08$ & $(0.41)$ \\ \hline
        \end{tabular}
\newline
\vskip 0.3cm
The number of events observed by H1 obviously deviates from the
SM expectation and from the ZEUS measurement whilst
the MC expectation after all cuts is in good agreement between 
both experiments. Higher statistics (HERA2) and smaller
uncertainties on the estimation of the dominant $W$-boson background 
(NLO calculation) will be needed to resolve this puzzle.

After applying further cuts on the isolated lepton events the final state
signature of a high $p_t^X$ jet and an isolated lepton has been used 
to test anomalous single production of top quarks $ep \rightarrow e t X$
where $t$ decays semi-leptonically. This process probes new physics at the
$ut\gamma$ vertex. Limits on the anomalous magnetic
coupling $\kappa_{ut\gamma}$ obtained by H1 and ZEUS \cite{ref:hera_top} are shown in fig.~\ref{pl:squark+ano}~(b) where H1
included also the hadronic decay channel $t\rightarrow b \rm \, jet \, jet$.
For comparison, 
also limits on $\kappa_{\gamma}$ and $v_Z$ (the anomalous vector coupling between $Z$, top and light quark) obtained by ALEPH and CDF are shown.

\section*{Summary}
Searches for compositeness, excited fermions,
quantum gravity effects, leptoquarks, and
\rpv SUSY have been performed and
no significant deviation from the SM was found at HERA1 by either of the
experiments H1 and ZEUS.
As presented, data taken at HERA1 have a big exclusion power 
for physics beyond the SM.
Currently the HERA accelerator is being upgraded 
to provide ten times higher integrated
luminosities and, in addition, longitudinally polarized $e^\pm$ beams.
Exploring the large discovery potential of HERA2 with the 
recently upgraded H1 and ZEUS 
detectors will certainly be interesting and help to resolve the puzzle
of the isolated lepton events.
An exciting time for searches to come... 

\section*{Acknowledgements}
I would like to thank my colleagues from H1 and ZEUS for their help
in preparing this talk.

%\newpage


\section*{References}
\begin{thebibliography}{99}

\bibitem{ref:nc_h1}
 { H1 Collaboration, ICHEP 2000 contributed paper, abstract~975.}

\bibitem{ref:nc_zeus}
   ZEUS Collaboration, ICHEP 2000 contributed paper, abstracts~1038, 1049. 

\bibitem{ref:osaka_ci}
 {H1 Collaboration, ICHEP 2000 contributed paper, abstract~952.}

\bibitem{ref:led} {N. Arkani-Hamed, S. Dimopolous and G. Dvali, Phys. Lett. B 429 (1998) 263 and Phys. D 59 (1999) 086004.}

\bibitem{ref:lq97_h1}
C.~Adloff {\it et al.}  [H1 Collaboration],
%``A search for leptoquark bosons and lepton flavor violation in e+ p  collisions at HERA,''
Eur.\ Phys.\ J.\ C {\bf 11}, 447 (1999)
[Erratum-ibid.\ C {\bf 14}, 553 (1999)]
[hep-ex/9907002].

%LQ e+p
\bibitem{ref:lq_h1_eplus}
 H1 Collaboration, ICHEP 2000 contributed paper, abstract~953. 

\bibitem{ref:dis97_zeus}
J.~Breitweg {\it et al.}  [ZEUS Collaboration],
%``Search for resonances decaying to e+ jet in e+ p interactions at HERA,''
Eur.\ Phys.\ J.\ C {\bf 16}, 253 (2000)
[hep-ex/0002038].


\bibitem{ref:brw}
W.~Buchmuller, R.~Ruckl and D.~Wyler,
%``Leptoquarks In Lepton Quark Collisions,''
Phys.\ Lett.\ B {\bf 191}, 442 (1987)
[Erratum-ibid.\ B {\bf 448}, 320 (1987)].

%LQ e-p
\bibitem{ref:lq_h1_emin}
 {H1 Collaboration, ICHEP 2000 contributed paper, abstract~954.}

\bibitem{ref:lq_zeus}
 {ZEUS Collaboration, contributed paper 552 to EPS 1999, Tampere, Finland,  July 1999.}
%Study of High Mass $e^-$-jet Systems in Electron-Proton Scattering at HERA 

%\bibitem{ref:lq_LEP}

%\bibitem{ref:lq_D0}

\bibitem{ref:lqind_h1}
{C.~Adloff {\it et al.}  [H1 Collaboration],
%``Search for compositeness, leptoquarks and large extra dimensions in e q  contact interactions at HERA,''
Phys.\ Lett.\ B {\bf 479}, 358 (2000)
[hep-ex/0003002].}

\bibitem{ref:exc_h1}
 {H1 Collaboration, ICHEP 2000 contributed paper, abstract~956;} \\
C.~Adloff {\it et al.}  [H1 Collaboration],
%``A search for excited fermions at HERA,''
Eur.\ Phys.\ J.\ C {\bf 17}, 567 (2000)
[hep-ex/0007035].

\bibitem{ref:exc_zeus}
   ZEUS Collaboration, ICHEP 2000 contributed paper, abstract~1040.

\bibitem{ref:exc}
F.~Boudjema, A.~Djouadi and J.~L.~Kneur,
%``Excited fermions at e+ e- and e P colliders,''
Z.\ Phys.\ C {\bf 57}, 425 (1993).

\bibitem{ref:exc_delphi}
% {DELPHI Collaboration, contributed paper 115 to EPS 1999, Tampere, Finland,  July 1999.}
DELPHI Collaboration, ICHEP 2000 contributed paper, abstract~192.


\bibitem{ref:dreiner}
J.~Butterworth and H.~Dreiner,
%``R parity violation at HERA,''
Nucl.\ Phys.\ B {\bf 397} (1993) 3
[hep-ph/9211204].

\bibitem{ref:rpsusy_h1}
C.~Adloff {\it et al.}  [H1 Collaboration],
%``Searches at HERA for squarks in R-parity violating supersymmetry,''
submitted to Eur.\ Phys.\ J.\ C, [hep-ex/0102050].

\bibitem{ref:rpsusy_zeus}
ZEUS Collaboration, ICHEP 2000 contributed paper, abstract~1042.

\bibitem{ref:isol}
C.~Adloff {\it et al.}  [H1 Collaboration],
%``Observation of events with an isolated high energy lepton and missing  transverse momentum at HERA,''
Eur.\ Phys.\ J.\ C {\bf 5}, 575 (1998)
[hep-ex/9806009]; \\
H1 Collaboration, ICHEP 2000 contributed paper, abstract~974.

\bibitem{ref:hera_top}
H1 Collaboration, ICHEP 2000 contributed paper, abstract~961; \\
ZEUS Collaboration, ICHEP 2000 contributed paper, abstract~1041.

\end{thebibliography}
\end{document}